\documentclass[a4paper,11pt]{article}
\pdfoutput=1
\usepackage{jheppub}
\usepackage{graphicx}
\usepackage[figureszright]{rotating}
\usepackage{bm,amsmath,amssymb}
\usepackage[mathscr]{eucal}
\usepackage{multirow}
\usepackage{xcolor}
\usepackage{mathrsfs}
\usepackage{mathtools}
\usepackage{slashed}
\usepackage{graphics}
\usepackage{graphicx}
\usepackage{subfigure}
\usepackage{dsfont}
\usepackage{longtable}
\usepackage{bbm} 
\usepackage{float}
\usepackage{tcolorbox}
\usepackage{lipsum}
\usepackage{cancel}
\usepackage{enumerate}
\usepackage{appendix}
\usepackage{soul}
\definecolor{lcolor}{rgb}{0.,0.0,0.}
\definecolor{citcolor}{rgb}{0,0.,0.5}

\def\bs{\boldsymbol} 
\def\del{\partial}

\newcommand{\eqn}[1]{Eq.~\eqref{#1}}
\newcommand{\eqsn}[1]{Eqs.~\eqref{#1}}

\newcommand{\nn}{\nonumber\\ }

\def\be{\begin{eqnarray*}}
\def\ee{\end{eqnarray*}}
\def\beq{\begin{eqnarray}}
\def\eeq{\end{eqnarray}}

\newcommand{\bea}{\beq \begin{aligned}}
\newcommand{\eea}{\end{aligned}\eeq}

\newcommand{\cO}{{\cal O}}

\newcommand{\bgamma}{{\boldsymbol \gamma}}
\newcommand{\bSigma}{{\boldsymbol \Sigma}}

\newcommand{\pT}{{p_T}}
\newcommand{\Qc}{S}
\newcommand{\Oc}{\mathcal{O}}

\newcommand{\rme}{{\rm e}}
\newcommand{\rmd}{{\rm d}}

\newcommand{\med}{{\rm med}}

\title{Energy–energy correlators and color decoherence from a generating functional}

\author[d]{Yacine Mehtar-Tani}
\emailAdd{mehtartani@bnl.gov}
\affiliation[d]{Physics Department, Brookhaven National Laboratory, Upton, NY 11973, USA}

\abstract{Using jet calculus at leading-logarithmic accuracy, we develop a generating-functional approach to describe the hard-collinear sector of jets propagating through the quark–gluon plasma. This framework yields coupled evolution equations for the jet nuclear modification factor and energy–energy correlators (EECs). We first construct the EEC evolution in the fully incoherent regime, and then derive evolution equations that account for the transition from the coherent to the incoherent limit in the large-$N_c$ limit. Furthermore, we show that combining the EEC with the jet nuclear modification factor provides complementary sensitivity to the overall jet suppression and to medium-induced modifications of the jet's internal structure. This framework therefore offers a way to constrain the magnitude and flavor dependence of jet energy loss, together with the medium resolution angle governing color decoherence.}


\begin{document}
\maketitle
\newpage 

\section{Introduction}

The discovery of jet quenching at RHIC in the early 2000s opened a new era in the study of QCD at high density, establishing jets as powerful probes of the quark--gluon plasma (QGP). Over the past two decades, experiments at RHIC and the LHC have accumulated a wealth of measurements of how jets are modified as they propagate through hot QCD matter~\cite{Connors:2017ptx,Apolinario:2022vzg,Apolinario:2024equ,STAR:2020ejj,STAR:2021kjt,ALargeIonColliderExperiment:2021mqf,ATLAS:2022vii,CMS:2024zjn}; see Ref.~\cite{Cunqueiro:2021wls} for a recent review of jet measurements. In parallel, substantial theoretical progress has extended the original picture of radiative energy loss by a leading parton~\cite{Gyulassy:1990ye,Wang:1991xy,Wang:1992qdg,Kovner:2003zj,Baier:1996kr,Baier:1996sk,Zakharov:1996fv,Zakharov:1997uu} to the evolution of full jets and their multi-parton structure~\cite{Mehtar-Tani:2010ebp,Mehtar-Tani:2011hma,Mehtar-Tani:2012mfa,Casalderrey-Solana:2011ule,Blaizot:2014bha,Iancu:2000hn,Caucal:2018dla,Caucal:2019uvr,Mehtar-Tani:2017web,Vaidya:2026yfa,Singh:2024vwb,Mehtar-Tani:2024smp,Mehtar-Tani:2025xxd,Caucal:2026lvn,Caucal:2026dsq,Caucal:2021bae,Abreu:2024wka,Arnold:2008zu,Arnold:2020uzm,Feal:2019xfl,Andres:2020vxs}; see also the recent review Ref.~\cite{Mehtar-Tani:2025rty} and references therein. More recently, effective-field-theory approaches have provided a systematic treatment of the underlying interference effects to all orders~\cite{Mehtar-Tani:2024smp,Mehtar-Tani:2025xxd,Vaidya:2026yfa,Caucal:2026dsq,Caucal:2026lvn}.

A central insight emerging from these developments is that a jet is a multi-parton system whose constituents do not necessarily interact with the medium independently. Depending on their angular separation relative to the resolving power of the medium, multiple partons may instead propagate as a single coherent color charge~\cite{Mehtar-Tani:2010ebp,Mehtar-Tani:2011hma,Mehtar-Tani:2011vlz,Casalderrey-Solana:2011ule,Mehtar-Tani:2011lic,Mehtar-Tani:2012mfa,Casalderrey-Solana:2012evi}. The transition from coherent to resolved energy loss is therefore governed by a characteristic angular scale, directly linking the modification of jet substructure to the resolving power of the QGP. Determining this scale experimentally, and thereby establishing to what extent the medium resolves the internal structure of jets, has become a central challenge in jet-quenching phenomenology~\cite{Mehtar-Tani:2016aco,Hulcher:2017cpt,Caucal:2018dla,Casalderrey-Solana:2018wrw,Caucal:2019uvr,Caucal:2021cfb,Pablos:2022mrx,Cunqueiro:2023vxl,Casalderrey-Solana:2019ubu,Mehtar-Tani:2021fud,Takacs:2021bpv,Attems:2022otp}.

Moreover, jet-substructure observables have attracted considerable phenomenological interest in recent years~\cite{STAR:2020ejj,STAR:2021kjt,ALargeIonColliderExperiment:2021mqf,ATLAS:2022vii,CMS:2024zjn}. Among these, the energy--energy correlator (EEC) has emerged as a particularly promising probe of medium modifications in the hard-collinear sector, as its energy weighting suppresses sensitivity to soft physics while retaining direct sensitivity to the angular structure of the jet~\cite{Andres:2022ovj,Andres:2023xwr,Andres:2024ksi,Andres:2024hdd,Barata:2023bhh,Barata:2025fzd,Yang:2023dwc,Singh:2024vwb}.

A major objective of jet-substructure studies has therefore been to determine the transverse resolution scale of the QGP. Until recently, however, color-coherence effects were primarily studied within analytic approaches and were not explicitly incorporated into many Monte Carlo event generators~\cite{Zapp:2013vla,Zapp:2026cqf,Caucal:2018ofz,Casalderrey-Solana:2014bpa,Chen:2017zte,Schenke:2009vr,Armesto:2009fj,JETSCAPE:2017eso}. They are now included, through different implementations, in frameworks such as the Hybrid model~\cite{Hulcher:2017cpt}, JetMed~\cite{Caucal:2018ofz}, and JEWEL~\cite{Zapp:2026cqf}. At the same time, an important alternative interpretation has emerged: several features of the observed jet modifications may arise predominantly from energy-loss-induced changes in the relative fractions of quark- and gluon-initiated jets, rather than from substantial medium modifications of the internal structure of individual jets~\cite{Spousta:2015fca,Qiu:2019sfj}. This effect, commonly referred to as selection bias~\cite{CMS:2020plq,Pradhan:2025tla}, can therefore mimic genuine medium-induced modifications of jet substructure and must be disentangled from them.

The extent to which a jet loses energy coherently, as a single color charge, or incoherently, through independently resolved constituents, therefore remains an open question. Recent analyses have argued that existing data already provide evidence for a finite medium-resolution scale~\cite{Kudinoor:2025gao,Pablos:2025cli,Mehtar-Tani:2024jtd}. Establishing such evidence within a systematically improvable framework, while disentangling genuine color-decoherence effects from modifications driven by flavor selection, is a central goal of the present work.

Further progress toward a quantitative understanding of jet modification requires analytic tools based on controlled approximations and systematic resummations. Such approaches can achieve precision in well-defined regions of jet phase space while being tailored to the observables of interest. Ultimately, they can also provide a theoretical foundation for the development of the next generation of Monte Carlo event generators.

In this work, we build on recent developments in jet-quenching theory based on the systematic resummation of collinear and soft logarithms~\cite{Mehtar-Tani:2017web,Mehtar-Tani:2017ypq,Mehtar-Tani:2024smp,Mehtar-Tani:2025xxd,Caucal:2026lvn,Vaidya:2026yfa}. These ideas have previously been applied to inclusive jet production, yielding successful descriptions of the jet nuclear modification factor at RHIC and the LHC~\cite{Mehtar-Tani:2017web,Mehtar-Tani:2021fud,Pablos:2025cli}. Here, we extend this framework to jet substructure by deriving the medium-modified energy--energy correlator (EEC) within a generating-functional formalism at leading-logarithmic (LL) accuracy.

In section~\ref{sec:eec_vac}, we begin by deriving the vacuum EEC at LL accuracy within the generating-functional approach. Then, in section~\ref{sec:eec_med}, we generalize the formalism to jets propagating through a medium, assuming a separation between the hard-collinear modes governing the energy-weighted correlator and the soft scales associated with interactions with the QGP. Within this factorized picture, the medium dependence is encoded in energy-loss distributions that provide the initial conditions for the subsequent collinear evolution. The resulting framework yields a unified description of the inclusive jet nuclear modification factor and the EEC distribution.

We first derive the evolution equations in the independent-energy-loss regime, where the angular separation between hard-collinear partons exceeds the coherence angle $\theta_c$. In this regime, the corresponding subjets are resolved by the medium and lose energy independently. This limit provides a natural starting point for incorporating color-coherence effects and describing the transition between coherent and resolved energy loss across the characteristic angular scale $\theta_c$.

In section~\ref{sec:eec_med_coh}, we show that, at LL accuracy, at most one splitting can probe the transition between coherent and decoherent energy loss. The associated interference effects are encoded through the energy loss of a color antenna, while subsequent emissions lie parametrically in either the fully coherent or independently resolved regime. Configurations in which multiple emissions simultaneously probe the transition contribute only beyond LL accuracy. We find that the medium modification of the EEC can be encoded in a medium-modified anomalous dimension whose angular dependence is governed by the dipole quenching factor. Our main results, given in Eqs.~(\ref{eq:gamma-coh}) and~(\ref{eq:EEC-sol-diff-med-2}), resum this angular-dependent evolution and predict a suppression of the EEC at large angles, above $\theta_c$, accompanied by an enhancement at small angles as a consequence of probability conservation.

Finally, in section~\ref{sec:eec_pheno}, we confront the framework with the CMS EEC data~\cite{CMS:2025ydi} for $R=0.4$ jets through a joint analysis with the ATLAS jet $R_{AA}$~\cite{ATLAS:2018gwx}, which allows the quark and gluon quenching factors to be constrained simultaneously. A detailed phenomenological analysis is presented in a companion Letter~\cite{Bossi:2026}. Since the present framework describes the hard-collinear core of the jet and does not include soft dynamics expected to become increasingly important at larger angles, we restrict the comparison to $\chi\lesssim 0.1$, where color-coherence effects are expected to dominate. In this region, the framework provides a satisfactory description of the characteristic enhancement of the PbPb-to-$pp$ EEC ratio at small angles followed by a suppression at larger angles, arising from the interplay between flavor-selection and color-decoherence effects. The subsequent rise of the ratio at larger angles, where the EEC itself rapidly decreases, lies outside the regime considered here and may receive important contributions from medium-induced soft radiation and medium response. By focusing on the hard-collinear region and simultaneously constraining the quenching factors with the jet $R_{AA}$, we find that neither the fully coherent nor the fully decoherent limit provides a consistent description of the data. Instead, the smooth transition encoded by the antenna quenching factor is favored, allowing us to extract the characteristic coherence angle $\theta_c$ of the medium. This analysis provides a first step toward a systematically improvable determination of the medium resolution scale, with higher logarithmic accuracy left for future work.

\section{EEC from Generating functional in vacuum }\label{sec:eec_vac}
In this work, we study the energy–energy correlator (EEC) measured within jets reconstructed with radius parameter $R$. The unnormalized differential EEC is defined as
\begin{align}
\frac{\mathrm{d}\Sigma}{\mathrm{d}p_T \mathrm{d}\chi}
&=
\sum_X
\int \mathrm{d}\Phi_X\, 
\frac{\mathrm{d}\sigma_X}
{\mathrm{d}p_T\mathrm{d} \Phi_X}
\sum_{\substack{i,j\in {\rm Jet}}}
\frac{p_{T,i}p_{T,j}}{p_T^2}\,
\delta\left(\chi-\theta_{ij}\right)\,,
\label{eq:eec-definition}
\end{align} 
where $\theta_{ij}$ denotes the measured angle between the particle pair, adopting the small-angle approximation $\theta_{ij}\ll 1$ and excluding self-correlations, $i\neq j$. Here, $p_T$ denotes the transverse momentum of the reconstructed jet, while $p_{T,i}$ and $p_{T,j}$ are the transverse momenta of its constituents. The quantity $\mathrm{d}\Phi_X$ denotes the phase-space measure for the final-state configuration $X$ within the jet, and $\mathrm{d}\sigma_X$ is the corresponding jet-production cross section. The sums run over all final-state configurations and over all distinct pairs of constituents belonging to the jet. In what follows we normalize the EEC by the jet cross section $\rmd\sigma_{\rm jet}/\rmd p_T$. 

\subsection{Jet calculus at leading logarithm accuracy }
Consider a generating functional that encodes jet fragmentation probabilities.   To construct the generating functional, we introduce an auxiliary probing function $u(k)$ associated with each final-state parton of momentum $k$. The generating functional is defined such that each parton in a given final state contributes one factor of $u(k)$. Functional derivatives with respect to $u(k)$ therefore generate the corresponding multiparton distributions, while specific choices of $u(k)$ can be used to construct the observable of interest. We have
\begin{align}\label{eq:GF-def}
Z[p,u]
&= P_1(p;k_1) u(k_1)+ P_2(p;k_1,k_2)\, u(k_1)u(k_2) +\cdots \nn
  &= \sum_{n=1}^{\infty}
  \int d\Phi_n\,P_n(p;\{k_i\})
  \prod_{i=1}^{n}u(k_i)\,,
  \end{align}
  where $P_n(p;{k_i})$ denotes the exclusive probability density for a hard parton of four-momentum $p$ to fragment into $n$ partons with four-momenta ${k_i}\equiv(k_1,k_2,\ldots,k_n)$, and $d\Phi_n$ denotes the corresponding $n$-particle phase-space measure.

Probability conservation implies
\begin{align}\label{eq:GF-norm}
Z[p,u=1]=1\,,
\end{align}
and we also have the constraint $Z[p=0]=1$. 

Let us define the cumulative EEC distribution for a given parton flavor (quark or gluon), normalized to the differential jet cross section, as

\begin{align}\label{eq:EEC-def}
 \Sigma_{q,g}(R/\chi)=\frac{1}{2}\left. \int_{1,2} \frac{E_1 E_2}{E^2} \Theta(\chi-\theta_{12}) \frac{\delta^2Z_{q,g}[p,u] }{\delta u(k_1)\delta u(k_2)}\right|_{u=1}
\end{align}
where $R$ is the jet opening angle. 

One can easily derive the evolution equation for the EEC from that of the generating functional \cite{Dokshitzer:1991wu,Dasgupta:2014yra}
\begin{align}\label{eq:GF-evol-q}
    \frac{\del Z_q[p,u]}{\del \ln R} &= \frac{\alpha_s}{\pi} \int \rmd z \,  p_{qq}(z) \left(  Z_q[zp,u]Z_g[(1-z)p,u] -Z_q[p,u]\right)\,,\nn
\end{align}
for the quark and similarly for the gluon 
\begin{align}\label{eq:GF-evol-g}
    \frac{\del Z_g[p,u]}{\del \ln R} &= \frac{\alpha_s}{\pi} \int \rmd z \,  p_{gg}(z) \left(  Z_g[zp,u]Z_g[(1-z)p,u] -Z_g[p,u]\right)\,\nn
    &+\frac{\alpha_s}{\pi} \int \rmd z \,  p_{qg}(z) \left(  Z_q[zp,u]Z_q[(1-z)p,u] -Z_g[p,u]\right)\,,\nn
\end{align}
where $p_{ij}(z)$ are the unregularized Altarelli-Parisi splitting functions \cite{Altarelli:1977zs}. 
The first term describes a real collinear splitting, in which the daughter partons carry momentum fractions $z$ and $1-z$, respectively. The second term is the corresponding virtual contribution required by probability conservation.
The unregularized real Altarelli--Parisi splitting functions are
\begin{align}\label{eq:AP-def}
p_{qq}(z)
&=
C_F \frac{1+z^2}{1-z}\,,\nn
p_{gg}(z)
&=
C_A
\left[
\frac{z}{1-z}
+
\frac{1-z}{z}
+
z(1-z)
\right] 
\,,\nn
\
p_{qg}(z)
&=
n_f T_R\left[z^2+(1-z)^2\right]\,,
\end{align}
with
\begin{align}
C_F=\frac{N_c^2-1}{2N_c}\,,
\qquad
C_A=N_c\,,
\qquad
T_R=\frac{1}{2}\,.
\end{align}
The $g\to q\bar q$ contribution is summed over the $n_f$ active quark flavors.

\subsection{Energy--energy correlators in vacuum }

Using \eqn{eq:EEC-def} together with \eqn{eq:GF-evol-q}, we readily obtain the contribution in which the two functional derivatives with respect to $u$ act on the same $Z$ factor,

\begin{align}
    Z_g \frac{\delta^2 Z_q}{\delta^2 u}+Z_q \frac{\delta^2 Z_g}{\delta^2 u} & \to \frac{\alpha_s}{2\pi} \int_0^1 \rmd z \,  \biggl[ z^2 p_{qq}(z) \Sigma_q (R/\chi) + (1-z)^2 p_{qq}(z) \Sigma_g(R/\chi) \biggr]\,,\nn
\end{align}
and, for the contribution in which each $Z$ factor receives one functional derivative,

\begin{align}
    \frac{\delta Z_q}{\delta u} \frac{\delta Z_g}{\delta u} & \to \frac{\alpha_s}{\pi} \int_0^1 \rmd z \,  \biggl[ p_{qq}(z)  \int_{1,2} \frac{E_1 E_2}{E^2} \Theta(\chi-\theta_{12}) \left. \frac{\delta Z_q[zp,u] }{\delta u(k_1)}\right|_{u=1} \left. \frac{\delta Z_g[(1-z)p,u] }{\delta u(k_2)}\right|_{u=1} \biggr]\,\nn
    &= \frac{\alpha_s}{2\pi}  \Theta(\chi-\theta_{12})  \int_0^1 \rmd z \,  z(1-z) p_{qq}(z)   \,,
\end{align}
and similarly for the other channels. 

Putting things together and integrating over $z$, we obtain 
\begin{align}\label{eq:EEC-diff}
    \frac{\del \Sigma_i(R/\chi) }{\del \ln R } = -\frac{\alpha_s}{2\pi} \,\Sigma_j(R/\chi)\,\gamma_{ji}(3)  + S_{i}(R/\chi)\,,
\end{align}
  where the anomalous dimensions are given by
  \begin{align}\label{eq:gamma-def}
\gamma_{qq}(3)&=-  2C_F  \int_0^1 \rmd z 
(z^2-1)\frac{1+z^2}{1-z}= \frac{25}{6} C_F\,,\nn
\gamma_{gq}(3)&=- 2 C_F  \int_0^1 \rmd z 
(1-z)^2\frac{1+z^2}{1-z}= - \frac{7}{6} C_F\,,\nn
\gamma_{gg}(3) &=
- 2 C_A  \int_0^1 \rmd z (z^2+(1-z)^2-1)
\left[
\frac{z}{1-z}
+
\frac{1-z}{z}
+
z(1-z)
\right] \nn
& + 2 n_f T_R \ \int_0^1 \rmd z [z^2+(1-z)^2]= \frac{14}{5}C_A +\frac{2}{3}n_f\,,\nn
\gamma_{qg}(3) &= - 
 2 T_R n_f  \int_0^1 \rmd z (z^2+(1-z)^2) \left[z^2+(1-z)^2\right] =-\frac{7}{15} n_f \,,
\end{align}
and the source term reads
\begin{align}
   S_{i}(R/\chi) &= \frac{\alpha_s}{2\pi}  \Theta(\chi-R) \left( \gamma_{qi}(3)+\gamma_{gi}(3)\right) \,.
\end{align}
The anomalous dimension in Eq.~\eqref{eq:gamma-def} coincides with the third Mellin moment of the regularized Altarelli--Parisi splitting functions \cite{Dixon:2019uzg}.
\beq 
\gamma(3) = - \int_0^1 \rmd z \, z^{2}  p_{\rm reg}(z)\,.
\eeq
The EEC evolution equation \eqn{eq:EEC-diff} is an inhomogeneous differential equation that can be readily solved, yielding, in matrix notation,
\begin{align}\label{eq:EEC-sol-cumul}
    \bSigma(R/\chi) &=\bSigma(R_0/\chi) \exp\left[- \frac{\alpha_s}{2\pi} \bgamma(3) \ln (R/R_0)\right] \nn&-\frac{\alpha_s}{2\pi}  \int_{R_0}^R \frac{\rmd \theta}{\theta} (1,1)\cdot \bgamma(3)\cdot \exp\left[- \frac{\alpha_s}{2\pi} \bgamma(3) \ln (R/\theta)\right] \, \Theta(\chi-\theta)\,,
\end{align}
where the contraction with $(1,1)$ implements the sum over final-state parton flavors.

The second term on the r.h.s. of \eqn{eq:EEC-sol-cumul} corresponds to the self-correlation contribution to the EEC, arising when the sum over particle pairs $i,j$ includes the case $i=j$, which is normalized to unity by virtue of the sum rule $\sum_i (E_i)^2=1$.

Neglecting the first term if we assume $R_0 < \chi < R$ we simply get 
\begin{align}\label{eq:EEC-sol-cum}
    \bSigma(R/\chi) &=\frac{\alpha_s}{2\pi} (1,1)\cdot \bgamma(3)\cdot \int_{0}^\chi \frac{\rmd \theta}{\theta} \exp\left[- \frac{\alpha_s}{2\pi} \bgamma(3) \ln (R/\theta)\right] \,\nn
    &=(1,1)\cdot\exp\left[- \frac{\alpha_s}{2\pi} \bgamma(3) \ln (R/\chi)\right] \,.
\end{align}
The differential distribution is then given by \cite{Dixon:2019uzg}

\begin{align}\label{eq:EEC-sol-diff}
   \frac{\del\bSigma(R/\chi) }{\del \ln \chi}  = \frac{\alpha_s}{2\pi} \cdot(1,1) \cdot  \bgamma(3)\cdot  \exp\left[- \frac{\alpha_s}{2\pi} \bgamma(3) \ln (R/\chi)\right]
\end{align}
where it is understood that
\begin{align}
\boldsymbol{\bSigma}
\equiv
\begin{pmatrix}
\Sigma_q \,,\,  \Sigma_g
\end{pmatrix},
\end{align}
and 
\begin{align}
\bgamma
\equiv
\begin{pmatrix}
\gamma_{qq}(3) \quad \gamma_{qg}(3)\\
\gamma_{gq}(3) \quad \gamma_{gg}(3)
\end{pmatrix},
\end{align}
Of course, this result can be easily generalized to running coupling case. 
\section{Jet-quenching effects on the EEC} \label{sec:eec_med}
The interaction of a high-energy jet with the medium is dominated by soft momentum transfers. We may therefore work at leading power in $\epsilon/\pT$, while resumming all powers of the parametrically enhanced combination $n\epsilon/\pT$, where $\epsilon\equiv E_{\rm loss}$ and $n$ characterizes the local steepness of the production spectrum. Within this approximation, the energy loss can be neglected in the hard-collinear fragmentation process and retained only through its effect on the hard production cross section of the initiating parton~\cite{Mehtar-Tani:2017web,Mehtar-Tani:2021fud,Mehtar-Tani:2024smp,Mehtar-Tani:2025xxd}.

This factorization is further supported by the separation of formation times between hard-collinear radiation and the characteristic medium time scales. In the angular regime $R\gtrsim\theta_c$, one has
\begin{align}
t_{\rm h.c.}\sim\frac{1}{\pT R^2}
\lesssim
\frac{1}{\pT\theta_c^2}
\ll L\,.
\end{align}
For example, taking $\pT=100~{\rm GeV}$, $\theta_c=0.1$, and $L=15~{\rm GeV}^{-1}\simeq3~{\rm fm}$ gives
\begin{align}
\frac{1}{\pT\theta_c^2}
\sim 1~{\rm GeV}^{-1}
\simeq0.2~{\rm fm}
\ll L\,.
\end{align}
The hard-collinear structure of the jet is therefore formed parametrically earlier than the medium-induced dynamics and can, to this accuracy, be factorized from the subsequent energy-loss process.

This separation considerably simplifies the treatment of multiple energy-loss processes. In particular, convolutions involving an arbitrary number of energy-loss probability distributions reduce to products of their Laplace transforms, evaluated at the conjugate variable $\nu=n/\pT$. Corrections to this approximation can, in principle, be incorporated systematically as power corrections in $\epsilon/\pT$, together with terms involving derivatives of the local spectral index $n$.

\subsection{Factorization of quenching weights}
To illustrate this factorization, consider the exclusive cross section for producing $m$ resolved final-state partons that are formed sufficiently early inside the medium, such that $t_{\rm h.c.}\ll L$. In vacuum, we have \cite{Mehtar-Tani:2024smp,Mehtar-Tani:2024mvl,Caucal:2026lvn}

\beq \label{eq:excl-vac}
        \frac{\rmd \sigma_\text{0,excl}}{\rmd k_1 \rmd k_2...\rmd k_N} && = \int \rmd p \, C(k_1,k_2,...,k_N|p) \,\delta(p-k_1-k_2-...-k_N )\,\frac{\rmd \sigma}{\rmd p },
\eeq
where we have used the shorthand notation
\beq
\rmd k_i \equiv  \frac{\rmd^2 k_{\perp i } \rmd k_i^+ }{2 (2\pi)^3 k^+_i}.
\eeq
\eqn{eq:excl-vac}  is modified in the medium as follows
\beq 
\frac{\rmd \sigma_\text{excl}}{\rmd k_1 \rmd k_2...\rmd k_N} && = \int \rmd p \prod_{i=1}^N\int \rmd \epsilon_i \, S(
 k_1,\cdots,k_N;\epsilon_1,\cdots,\epsilon_N) \times C(k_1+\epsilon_1,k_2+\epsilon_2,...,k_N+\epsilon_N|p) \nn
&&\times\delta(p-\epsilon-k_1-k_2-...-k_N )\,  \frac{\rmd \sigma}{\rmd p }\,, \nn
\eeq

where $\epsilon=\epsilon_1+\cdots+\epsilon_N$ is the total energy lost to the plasma by the initial parton of momentum $p$.

We first develop the framework in the independent energy-loss approximation, in which the final-state hard-collinear modes lose energy incoherently, allowing us to assume the factorization

\begin{align}
 S(
 k_1,\cdots,k_n;\epsilon_1,\cdots,\epsilon_n)
   \approx  \prod_{i=1}^N  P(\epsilon_i) \,.
\end{align}
This provides a convenient intermediate step toward incorporating interference effects that encode color coherence.

In the above expression it is understood that  the $\epsilon$'s appear only in the $k^+$ component of the 3-vectors. 
Shifting $p \to p+\epsilon = p+\sum \epsilon_i$ and neglecting $\epsilon_i$'s in the hard-collinear coefficients $C$, but not in the initial power spectrum $\rmd \sigma/\rmd p$, owing to the fact that multiple branchings in the vacuum do not generate any large powers of the fragment energies, the above expression simplifies as 
\beq 
\frac{\rmd \sigma_\text{excl}}{\rmd k_1 \rmd k_2...\rmd k_N} && \simeq  \int \rmd p  \left\{ \int \prod_{i=1}^N \rmd \epsilon_i P(\epsilon_i) \right\}  C(k_1,k_2,...,k_N|p) \nn
&&\times\delta(p-k_1-k_2-...-k_N )\,  \frac{\rmd \sigma(p+\epsilon)}{\rmd p }+\Oc(1/n) \nn
&& \simeq \left( \int_0^\infty  \rmd \epsilon P(\epsilon) \exp\left(-\frac{n\epsilon}{\pT}\right)\right)^N  \frac{\rmd \sigma_\text{0,excl}}{\rmd k_1 \rmd k_2...\rmd k_N} \nn
&& =  \, \,S(\nu=n/p)^N\,\,\frac{\rmd \sigma_\text{0,excl}}{\rmd k_1 \rmd k_2...\rmd k_N}.
\eeq
where
\beq\label{eq:LT_qw}
S(\nu) = \int_0^\infty  \rmd \epsilon P(\epsilon)\, \rme^{-\nu\epsilon}\,,
\eeq
is the Laplace transform of the energy-loss distribution, which, when evaluated at $\nu=n/p$, will be referred to as the quenching factor.

Color-coherence effects have recently been formalized in terms of collinear-soft functions involving correlators of Wilson lines associated with the final-state hard-collinear modes~\cite{Mehtar-Tani:2024smp,Mehtar-Tani:2025xxd}. In the present discussion, we assume that these soft functions factorize, as appropriate when the angular separations between hard-collinear modes are much larger than the medium coherence angle, $\theta_c$, such that they lose energy independently. This approximation leads to a nonlinear DGLAP evolution that resums large logarithms of $R/\theta_c$ and has been successfully applied in recent phenomenological studies~\cite{Mehtar-Tani:2021fud,Mehtar-Tani:2024jtd,Pablos:2025cli}. Here, we recover these evolution equations directly within the generating-functional approach. In the next section, we go beyond the independent-energy-loss approximation and incorporate the interference effects governing the transition between coherent and resolved energy loss at LL accuracy.

The first correction to the above result takes the form
\beq
N\Qc(p)^{N-1}
\frac{\del \Qc(p)}{\del (n/p)}
\sum_{i=1}^N
\frac{\del}{\del k_i}
\left(
\frac{\rmd \sigma_{\text{0,excl}}}
{\rmd k_1 \rmd k_2\cdots\rmd k_N}
\right).
\eeq
Thus, within this simple picture and at leading power in the energy loss, the exclusive spectrum is suppressed by a product of quenching factors, with one factor associated with each resolved parton. Corrections to this factorized form are sensitive to the detailed kinematics of the underlying $1\to N$ splitting process and enter at higher order in the power expansion.

It is useful to keep the Laplace variable $\nu$ explicit throughout the evolution and set $\nu=n/p_T$ only at the end of the calculation, where $n$ denotes the local spectral index of the initiating-parton production spectrum.

\subsection{Generating-functional approach to jet quenching }
It is now straightforward to generalize the generating-functional evolution equation \eqn{eq:GF-evol-q} to the case of a jet undergoing energy loss in a hot QCD medium. Within the independent-energy-loss approximation, the evolution equation retains the same form as in vacuum.  The main difference lies in the initial condition,
\beq
Z[p,u,\theta=0]=\Qc_0(\nu)\,u(p)\,,
\eeq
where $\Qc_0(\nu)$ denotes the quenching factor of a single parton. We emphasize that this approximation applies at angles larger than the medium resolution angle $\theta_c$, where the hard-collinear modes are independently resolved by the medium. An infrared separation scale between the hard-collinear and soft medium modes is also implicitly assumed. This scale enters the quenching factors and their evolution~\cite{Caucal:2026lvn}, but, as we shall see, does not directly affect the evolution of the EEC.

In the following, we suppress the explicit dependence on the Laplace variable $\nu$, since it does not participate in the QCD evolution of the jet. This also avoids confusion between the fixed Laplace argument, ultimately evaluated at $\nu=n/p_T$, and the momentum dependence dynamically generated by the collinear evolution. Indeed, the all-order QCD evolution generates a nontrivial dependence of the quenching factor on the momentum $p$ of the initiating parton. The resulting all-order quenching factor is obtained from the normalization condition

\beq
Z[p,u=1,\theta]=\Qc(p,\theta)\,.
\eeq
Unlike in vacuum, the generating functional for the measured jet sector is therefore not normalized to unity, but to the jet quenching factor. This does not imply a violation of probability conservation in the complete system; rather, probability is transferred outside the measured jet sector through energy loss to the medium. The evolution equation for $\Qc(p,\theta)$ follows directly by setting $u=\Qc_0(\nu)$ in \eqsn{eq:GF-evol-q} and \eqref{eq:GF-evol-g}. For the quark jet, for instance,  we have,

\begin{align}\label{eq:qw-evol-med}
    \frac{\del \Qc_q(p,\theta)}{\del \ln \theta} = \frac{\alpha_s}{\pi} \int_0^1 \rmd z \,  p_{qq}(z) \left[ \Qc_q(zp,\theta)\Qc_g((1-z)p,\theta) -\Qc_q(p,\theta)\right]\,,
\end{align}
and similarly for gluon jets. 
The infrared safety of the evolution equation is ensured by the boundary condition $S(p=0)=1$, which reflects the fact that a parton with vanishing initial energy has no energy to lose. 

\eqn{eq:qw-evol-med} was first proposed in \cite{Mehtar-Tani:2017web} and subsequently applied to phenomenological studies in \cite{Mehtar-Tani:2021fud,Mehtar-Tani:2024jtd,Pablos:2025cli}.
The quark and gluon quenching factors are related to the jet nuclear modification factor through
\beq
R_{AA}=f_q\,\Qc_q+f_g\,\Qc_g\,,
\eeq
where
\beq
f_q=
\left(\frac{\rmd\sigma_{\rm jet}^{pp}}{\rmd p_T}\right)^{-1}
\frac{\rmd\sigma_q^{pp}}{\rmd p_T}\,, \qquad f_g =
\left(\frac{\rmd\sigma_{\rm jet}^{pp}}{\rmd p_T}\right)^{-1}
\frac{\rmd\sigma_g^{pp}}{\rmd p_T}\,,
\eeq
denote the quark- and gluon-initiated jet fractions in $pp$ collisions, respectively, and satisfy $f_q+f_g=1$.
\begin{figure}
\begin{center}
\includegraphics[width=0.7\textwidth]{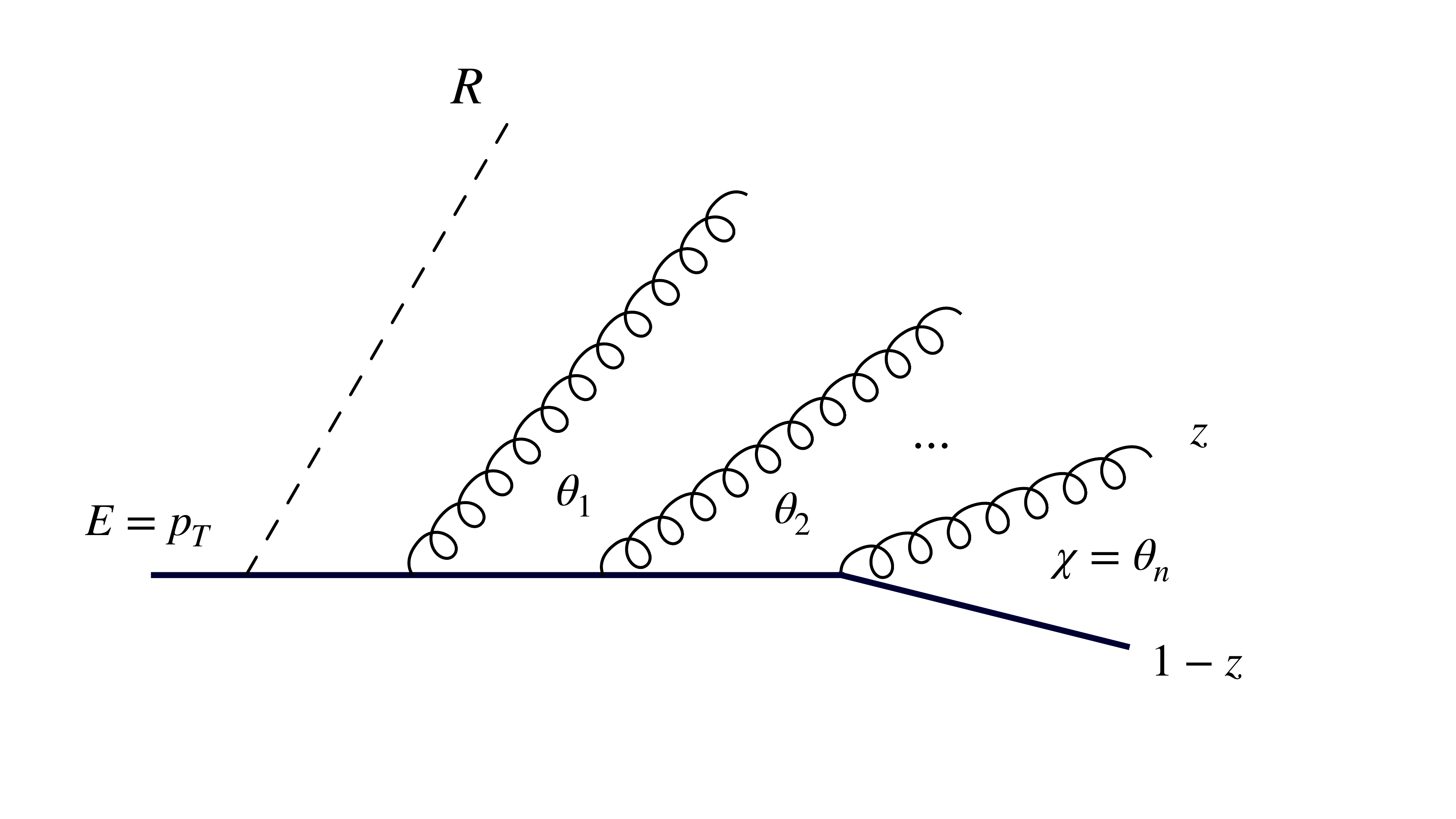} 
\end{center}
\caption{
Schematic representation of the vacuum contribution to the EEC in an
angular-ordered gluon cascade initiated by a jet of energy $E=p_T$ and
radius $R$. Successive emissions are strongly ordered in angle,
$R\gg\theta_1\gg\theta_2\gg\cdots\gg\theta_n$, while the last resolved
splitting, with energy fractions $z$ and $1-z$, fixes the measured EEC
angle, $\chi=\theta_n$.
 }
\label{fig:eec_pp}
\end{figure}
\subsection{Medium-modified energy-energy correlators }
\begin{figure}
\begin{center}
\includegraphics[width=0.8\textwidth]{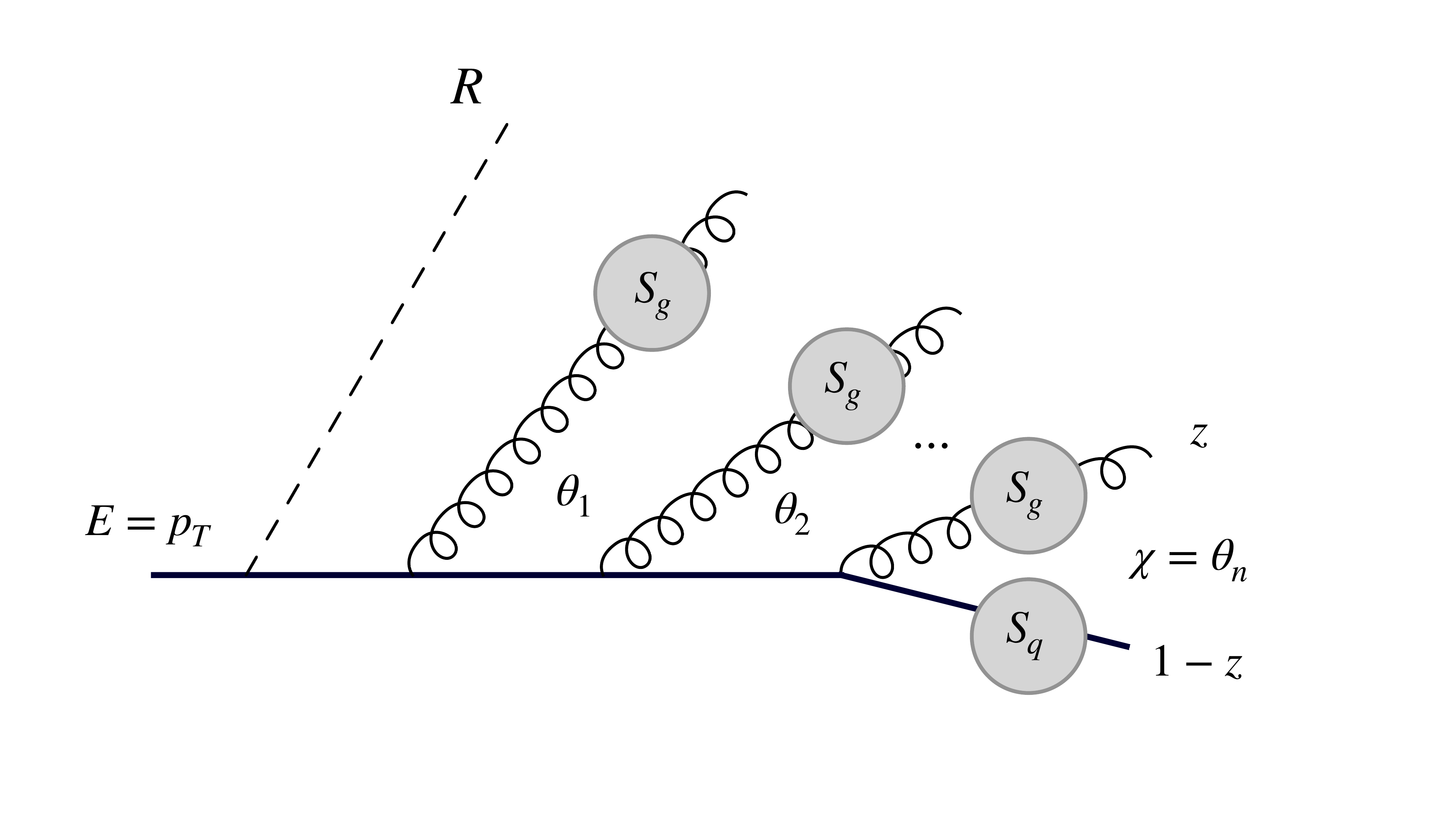} 
\end{center}
\caption{
Schematic representation of the medium-modified EEC in an
angular-ordered gluon cascade in the incoherent energy loss approximation. Successive emissions are strongly ordered,
$R\gg\theta_1\gg\theta_2\gg\cdots\gg\theta_n$, and the last resolved
splitting, with energy fractions $z$ and $1-z$, fixes the measured angle,
$\chi=\theta_n$. The gray disks denote the gluon quenching weights
$S_g$ associated with the color charges produced throughout the cascade. 
 }
\label{fig:eec_quenching}
\end{figure}

Let us now return to the EEC problem. We use essentially the same definition as in vacuum, \eqn{eq:EEC-def}, but with a different normalization,

\begin{align}\label{eq:EEC-def-med}
 \Sigma(R/\chi)=\frac{1}{2}\, \cdot\,\frac{1}{Z[u=1]}\left. \int_{1,2} \frac{E_1 E_2}{E^2} \Theta(\chi-\theta_{12}) \frac{\delta^2Z[p,u] }{\delta u(k_1)\delta u(k_2)}\right|_{u=1}\,,
\end{align}
and applying the same chain rules we find (omitting the flavor indices for simplicity): 
\begin{align}\label{eq:eec-evol-med-1}
\frac{\del}{\del \ln \theta}S(\theta)\, \Sigma(\theta/\chi) & = \frac{\alpha_s}{\pi} \int_0^1 \rmd z \,  p(z)  \left[ (z^2  + (1-z)^2)S(zp)S((1-z)zp)-  S(p)\right]\Sigma(\theta/\chi)\nn
&+\frac{\alpha_s}{\pi}  \Theta(\chi-\theta) \int_0^1 \rmd z z(1-z)S(z) S(zp)S((1-z)zp)\,.
\end{align}
Subtracting unity from $z^2+(1-z)^2$ reconstructs the infrared-safe combination $z^2+(1-z)^2-1$, thereby generating the quenching-factor evolution kernel appearing on the right-hand side of \eqn{eq:qw-evol-med}.
The l.h.s. can be rewritten as 
\beq 
\frac{\del}{\del \ln \theta}S(\theta)\, \Sigma(\theta/\chi)  =  \Sigma(\theta/\chi) \, \frac{\del}{\del \ln \theta}S(\theta)\,+S(\theta)\, \frac{\del}{\del \ln \theta}\, \Sigma(\theta/\chi)\,.
\eeq
Hence,
\begin{align}
\frac{\del}{\del \ln \theta}S(\theta)\, \Sigma(\theta/\chi) & = \frac{\alpha_s}{\pi} \int_0^1 \rmd z \,  p(z)  \left[ (z^2  + (1-z)^2-1)S(zp)S((1-z)zp)\right]\Sigma(\theta/\chi)\nn
&+ \frac{\alpha_s}{\pi} \int_0^1 \rmd z \,  p(z)  \left[ S(zp)S((1-z)zp)-S(p)\right] \Sigma(\theta/\chi)
\nn
&+\frac{\alpha_s}{\pi}  \Theta(\chi-\theta) \int_0^1 \rmd z z(1-z)p(z) S(zp)S((1-z)zp)\,,
\end{align}
and we see that the first term cancels exactly the second line in \eqn{eq:eec-evol-med-1}. As a result, we  obtain

\begin{align}\label{eq:eec-evol-med-2}
\, \frac{\del}{\del \ln \theta} \Sigma(\theta/\chi) & = \frac{\alpha_s}{\pi} \int_0^1 \rmd z \,  p(z)  (z^2  + (1-z)^2-1)S(zp)S((1-z)p)S^{-1}(p)\, \Sigma(\theta/\chi)\nn
&+\frac{\alpha_s}{\pi}  \Theta(\chi-\theta) \int_0^1 \rmd z z(1-z)p(z) S(zp)S((1-z)zp)S^{-1}(p)
\end{align}
from which we can read the medium modified anomalous dimension 
\beq 
\gamma_{\rm med}(3)  = - \int_0^1 \rmd z (z^2  + (1-z)^2-1)  S(zp)S((1-z)zp)S^{-1}(p)\,. 
\eeq
A diagrammatic depiction of the above evolution equation is given in figure~\ref{fig:eec_quenching}.

The quenching factor $S(\nu,p)\approx S(\nu)$ depends only weakly on the parton momentum $p$ through the evolution (cf.~\eqn{eq:bms}). Consequently, its $z$ dependence can be neglected at LL accuracy. For instance, in the purely gluonic case, this yields the simple relation between the vacuum and medium-modified anomalous dimensions:
\beq 
\gamma_{gg,\rm med}(3)  \simeq S_g(\nu)\, \gamma_{gg}(3) \,.
\eeq

In this approximation the solution is readily obtained from \eqn{eq:EEC-sol-diff}
\begin{align}\label{eq:EEC-sol-diff-med}
   \frac{\del\bSigma(R/\chi) }{\del \ln \chi}  = \frac{\alpha_s}{2\pi} (1,1)\cdot\bgamma_{\rm med}(3) \cdot \exp\left[- \frac{\alpha_s}{2\pi} \bgamma_{\rm med}(3) \ln (R/\chi)\right]\,.
\end{align}
The medium-modified anomalous dimension is explicitly given by the $2\times 2$ matrix 
\begin{align}
\bgamma_{\rm med}(3)
\equiv
\begin{pmatrix}\label{eq:gamma-incoh}
\frac{25}{6}C_F \,S_g  \quad -\frac{17}{15}n_f \, S_q^2/S_g \\
-\frac{7}{6}C_F \,S_g\quad \frac{14}{5}C_A S_g +\frac{2}{3}n_f 
\end{pmatrix},
\end{align}
Note that no additional quenching factor appears in the term proportional to $n_f$ in $\gamma_{gg}$. This term arises from the virtual quark-loop correction to the gluon channel and therefore carries the total gluon quenching factor $S_g$, which cancels against the corresponding total-charge quenching factor in the denominator. 

Setting $S=1$ in \eqn{eq:gamma-incoh}, we recover the vacuum anomalous dimension \eqn{eq:gamma-def}.

Color-coherence effects can be incorporated through an interpolation that reproduces the leading correction near the coherent limit. This correction is proportional to the dipole color-singlet quenching factor, which will be discussed in the next section. Away from the coherent regime, when $\theta\gg\theta_c$, the anomalous dimension smoothly approaches the fully incoherent limit given in \eqn{eq:gamma-incoh}.

\section{EEC and color decoherence  }\label{sec:eec_med_coh}
So far, we have discussed the EEC and jet quenching within a fully incoherent picture, which remains the default assumption in most Monte Carlo event generators. Color-coherence effects have recently been incorporated into several generators \cite{Caucal:2018ofz,Hulcher:2017cpt,Zapp:2026cqf}, and their phenomenological importance is increasingly recognized, particularly for the simultaneous description of multiple jet observables \cite{Kudinoor:2025gao}. 
\subsection{Factorization of dipoles at LL approximation  }
\begin{figure}
\begin{center}
\includegraphics[width=0.7\textwidth]{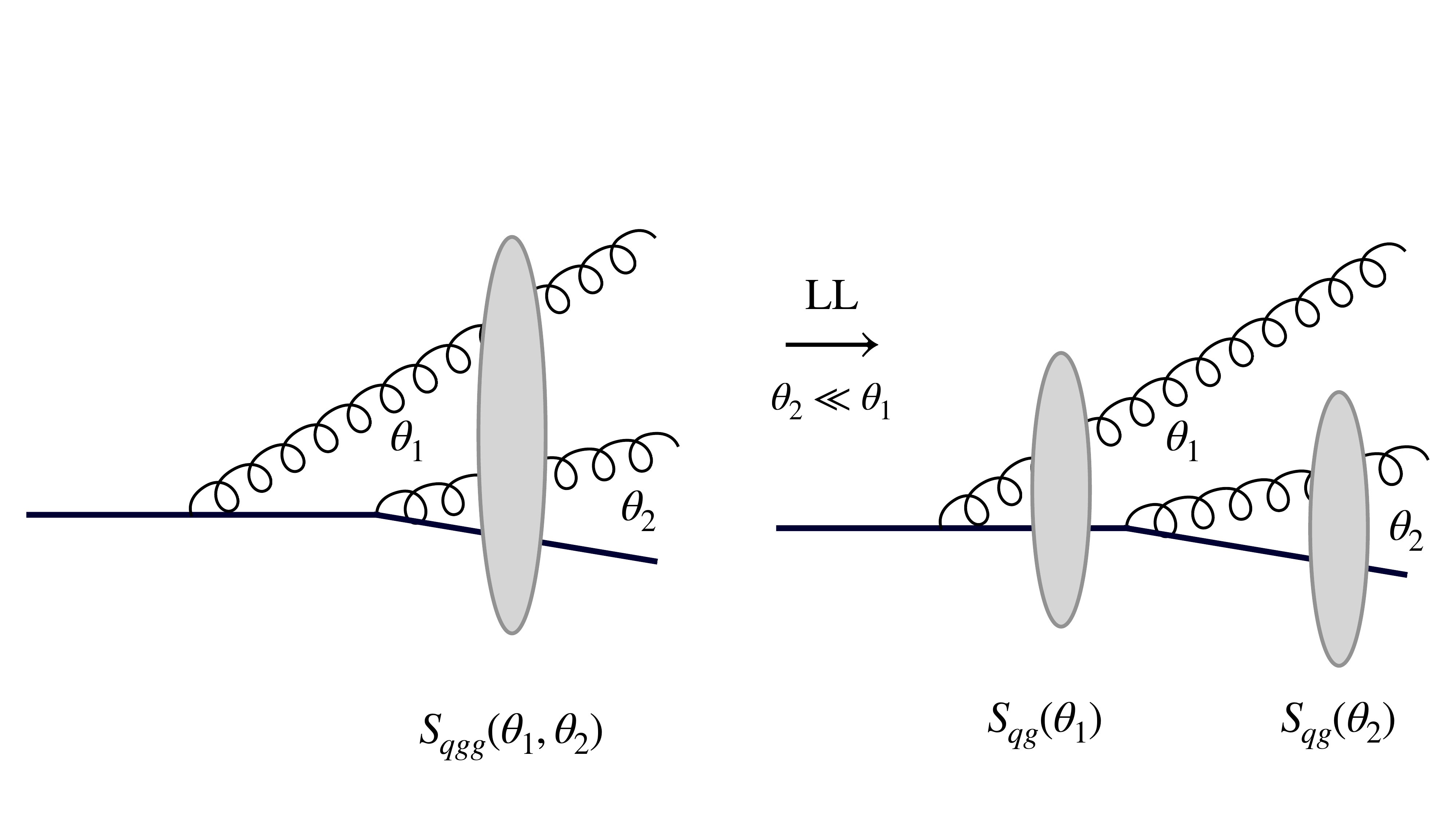} 
\end{center}
\caption{
Diagrammatic representation of the real-emission contribution to the EEC at NLO. In the strongly angular-ordered regime, $\theta_2\ll\theta_1$, the energy-loss correlator of the three-body color system factorizes as
$
S_{qgg}(\theta_1,\theta_2)
\simeq
S_{qg}(\theta_1)\,S_{qg}(\theta_2).
$
At leading-logarithmic accuracy, only one emission can probe the transition region, $\theta_i\sim\theta_c$. The other is parametrically separated from $\theta_c$ and therefore lies either in the fully coherent regime, $\theta_j\ll\theta_c$, or in the fully decoherent regime, $\theta_j\gg\theta_c$. Corrections to this factorized picture arise at NLL accuracy. 
 }
\label{fig:eec_fact}
\end{figure}

To illustrate the basic idea underlying the factorization of the color-decoherence transition that we shall construct to all orders, consider a configuration involving two successive collinear splittings, $q\to qg$ followed by $q\to qg$, occurring at angles $\theta_1$ and $\theta_2$, respectively, as depicted in figure~\ref{fig:eec_fact}. Although the collinear splitting probabilities factorize, the corresponding energy-loss distribution need not, due to soft-gluon interference among the final-state partons. At leading-logarithmic accuracy, however, successive emissions are strongly angular ordered, $\theta_1\gg\theta_2$. Consequently, if one of the emissions occurs near the transition angle $\theta_c$, the other necessarily lies parametrically far from the transition region, either in the fully coherent or in the fully decoherent regime. This observation leads to the factorization
\begin{align}\label{eq:fact-dip}
S_{qgg}(\theta_1,\theta_2)
\underset{\theta_2\ll\theta_1}{\simeq}
S_{qg}(\theta_1)\, S_{qg}(\theta_2)\, ,
\end{align}
up to power corrections in $\theta_2/\theta_1$.

Let us examine the two limiting configurations relevant to this factorization. First, consider $\theta_1\sim\theta_c$. Strong angular ordering then implies $\theta_2\ll\theta_c$, so that the second splitting is unresolved by the medium and the corresponding daughter system behaves as a single quark charge, $S_{qg}(\theta_2)\simeq S_q$. Hence,
\begin{align}
S_{qgg}(\theta_1,\theta_2)
\to
S_{qg}(\theta_1)\, S_q 
\qquad
\text{for}\qquad
\theta_1\sim\theta_c\, ,\quad \theta_2\ll\theta_1\, .
\end{align}

Conversely, if the second splitting occurs near the transition, $\theta_2\sim\theta_c$, angular ordering requires $\theta_1\gg\theta_c$. The first splitting is then fully resolved by the medium, such that the gluon produced at the larger angle quenches as an independent color charge. One therefore obtains
\begin{align}
S_{qgg}(\theta_1,\theta_2)
\to
S_g\, S_{qg}(\theta_2)
\qquad
\text{for}\qquad
\theta_2\sim\theta_c\, ,\quad \theta_1\gg\theta_2\, .
\end{align}
Here we have neglected the angular dependence of the single-charge quenching factors, assuming $S_{q,g}(\theta)\simeq S_{q,g}(R)$ for angles well inside the jet, $\theta\ll R$.

This factorization naturally generalizes to higher orders. Corrections first arise at NLL accuracy, where successive gluons can be emitted at comparable angles and may therefore simultaneously probe the transition region around $\theta_c$.

The final ingredient of our approach is the large-$N_c$ limit. In this limit, the $qg$ quenching factor further factorizes into the quenching factor associated with the total color charge and a color-singlet antenna, or dipole, quenching factor,
\begin{align}
S_{qg}(\theta) \simeq S_q(\theta)\, S_{\rm dip}(\theta)\,.
\end{align}
The angular dependence of $S_q$ is generated by quantum evolution \eqn{eq:bms}. 

This factorization cleanly separates the overall quenching of the parent color charge from the angular dependence associated with the medium resolution of the $qg$ system. In the following section, we recall the expression for the dipole quenching factor obtained in the harmonic-oscillator approximation in Ref.~\cite{Mehtar-Tani:2017ypq}. An illustration of the emergence of the dipole structure in the large-$N_c$ limit is shown in figure~\ref{fig:eec_dipoles}.

\subsection{The singlet antenna quenching factor  }
The key quantity entering our analysis is the antenna (dipole) quenching factor, first derived in Ref.~\cite{Mehtar-Tani:2017ypq}:
\begin{align}
\label{eq:antenna-qw}
&S_{\rm dip}(\theta_{12},\nu) =   S_1(\nu, L) \, S_2(\nu, L)\nn
 &- 2 \int_0^L \rmd t  \, S_1(\nu, L-t) \, S_2(\nu, L-t)  \Big[ 1- \Delta_\med(t) \Big]  \, \Gamma (\nu)\,,
\end{align}

where 
\beq 
\Gamma (\nu) = \int_0^\infty \rmd \omega \int \rmd \vartheta \frac{\rmd I}{\rmd \omega \rmd \vartheta} (1- \rme ^{-\nu \omega}) \Theta(\vartheta > R)\,.
\eeq
Here, $\Gamma(\nu,R)$ denotes the medium-induced radiation rate in Laplace space, evaluated in the independent multiple-soft-gluon-emission approximation~\cite{Mehtar-Tani:2025rty}. This description applies to a sufficiently extended medium and to soft gluon emissions with formation times much shorter than the medium length, $t_f\ll L$. The factor $\Theta(\vartheta>R)$ restricts the radiation to angles outside the jet cone, since only such emissions contribute directly to the out-of-cone energy loss. The corresponding quenching factor obeys
\beq\label{eq:rate-eq-1}
\frac{\partial S_1}{\partial t}
= -\Gamma(\nu)\, S_1\,.
\eeq
where $t=L$ at the end of the evolution.

The medium decoherence parameter $\Delta_{\rm med}(t)$~\cite{Mehtar-Tani:2010ebp,Mehtar-Tani:2011hma,Casalderrey-Solana:2011ule}, governs the transition between coherent and independently resolved energy loss. For a dense medium,
\begin{align}
\Delta_{\rm med}(t)
=
1-\exp\!\left[
-\frac{1}{12}\hat q\,\theta_{12}^2 t^3
\right],
\end{align}
which introduces the characteristic angle 
\beq\label{eq:theta_c}
\theta_c
=\sqrt{\frac{12}{\hat q L^3}}\,.
\eeq

Dipoles with $\theta_{12}\ll\theta_c$ remain color coherent and interact with the medium as a single color charge. Using the rate equation \eqref{eq:rate-eq-1}, 
one finds $S_{12}(\nu)\to 1$, corresponding in energy space to
\beq
S_{\rm dip}(\varepsilon)=\delta(\varepsilon)\,.
\eeq
In the opposite limit, $\theta_{12}\gg\theta_c$, the medium resolves the two constituent charges, which then lose energy independently. Consequently, the dipole energy-loss factor becomes
\beq
S_{\rm dip}(\nu)
\to
S_1(\nu)S_2(\nu)\, .
\eeq
In energy space, this product corresponds to the convolution of the two independent energy-loss distributions.
This behavior is analogous to that of the dipole $S$-matrix, $S(r_\perp)$, describing the scattering of a color dipole of fixed transverse size $r_\perp$ in the McLerran--Venugopalan model \cite{McLerran:1993ka,McLerran:1993ni}. For $r_\perp\ll1/Q_s$, where $Q_s$ is the nuclear saturation scale, color transparency implies $S(r_\perp)\to1$ \cite{Caucal:2026dsq}. Conversely, for $r_\perp\gg1/Q_s$, the dipole enters the strong-scattering regime and $S(r_\perp)\to0$.

The key observation underlying the incorporation of color-coherence effects into the EEC calculation of the previous section is that, at leading-logarithmic accuracy (LLA), the angles of successive splittings are strongly ordered, as illustrated in figure~\ref{fig:eec_pp}. Consequently, the transition from coherent to decoherent energy loss around the characteristic angle $\theta_c$ involves only one splitting in the angular-ordered sequence. Configurations in which two emissions simultaneously satisfy $\theta_1\sim\theta_2\sim\theta_c$ are not logarithmically enhanced and are therefore beyond LLA.
\begin{figure}
\begin{center}
\includegraphics[width=0.9\textwidth]{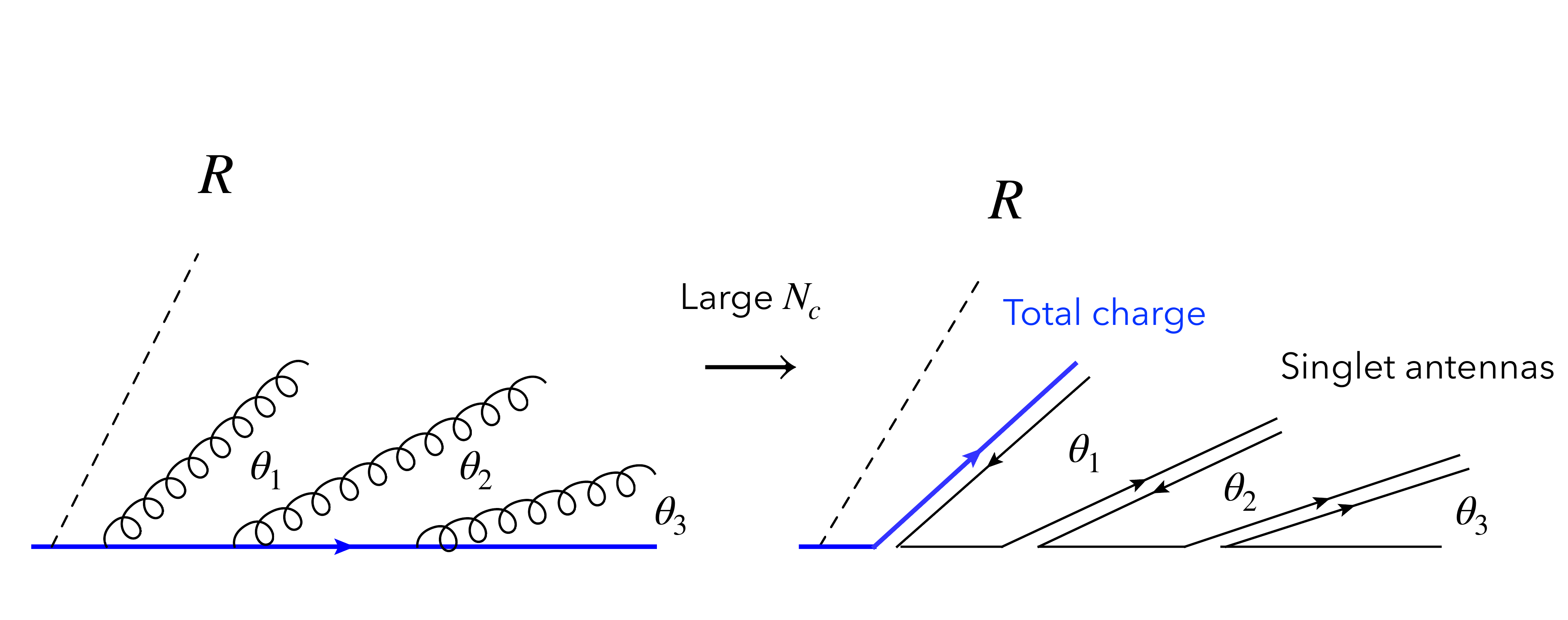} 
\end{center}
\caption{
Schematic representation of the large-$N_c$ color structure of an 
angular-ordered jet cascade. Successive splittings occur at strongly 
ordered angles, $R \gg \theta_1 \gg \theta_2 \gg \theta_3$. At 
leading-logarithmic accuracy, each subsequent, smaller-angle splitting 
is unresolved by the preceding antenna and can therefore be neglected 
in its quenching factor. The cascade can thus be reorganized into the 
total jet color charge (blue), which controls the overall quenching, 
and a sequence of color-singlet antennas whose dipole quenching factors 
encode the transition from coherent to decoherent energy loss.
 }
\label{fig:eec_dipoles}
\end{figure}

Hence, as discussed above, we incorporate color-coherence effects minimally by working in the large-$N_c$ limit (cf.~figure\ref{fig:eec_dipoles}) and making, for instance, the following replacement for a quark--gluon antenna in \eqn{eq:eec-evol-med-2}:
\beq
\Qc_g(zp,\theta)\Qc_q((1-z)p,\theta)
\,\longrightarrow\,
\Qc_q(p,\theta)\,S_{\rm dip}(p,z,\theta)\,.
\eeq

Here, we have used the factorization \eqn{eq:fact-dip}, which, upon neglecting higher-order corrections, reduces to \eqn{eq:antenna-qw}. Furthermore, if the quenching factors depend only weakly on the parton momentum, the $z$ dependence of $S_{\rm dip}$ may be neglected.

This modification of the incoherent picture is most transparent in the double-logarithmic approximation, in which the evolution of the quenching factor is given by \cite{Caucal:2026lvn,Caucal:2026dsq,Vaidya:2026yfa} 
\begin{align}\label{eq:bms}
    S_q(p) = S^{(0)}_q + \frac{\alpha_s C_F}{\pi} \int_{\mu_{\rm soft}/p}^1 \frac{\rmd z}{ z} \int_0^R \frac{\rmd \theta_{2} }{\theta_{2}}\left[ S_q(\theta_2,p) S_{\rm dip}(\theta_{12},p) -S_q(\theta_1) \right] \,.
\end{align}
where $S_q^{(0)}$ denotes the bare quenching factor and $\mu_{\rm soft}$ is an infrared scale that regulates the soft singularity and is related to medium-induced energy loss. As shown in \cite{Mehtar-Tani:2024mvl}, this scale is parametrically related to the power-law index $N$ of the jet production spectrum, defined by $\sigma_{pp}(p_T)\sim p_T^{-N}$, with $N\sim4\text{--}10$, through the relation $\mu_{\rm soft}/p_T\sim1/N\ll1$.

At NLO, for instance we have
\begin{align}
    S_q^{(1)} \approx S^{(0)}_q + \frac{\alpha_s C_F}{\pi} \ln N  \int_0^R \frac{\rmd \theta_{2} }{\theta_{2}}\left[ S^{(0)}_q(\theta_2) S^{(0)}_{\rm dip}(\theta_{12}) -S^{(0)}_q(\theta_1) \right]+\cO(\alpha_s^2) \,.
\end{align}
Ref.~\cite{Caucal:2026dsq} shows that the magnitude of these corrections remains relatively modest as a consequence of color-coherence effects. However, in the regime $p_T/N\gg\mu_{\rm soft}$, the available energy phase space generates a potentially large logarithm, $\ln(p_T/\mu_{\rm soft})$, leading to an additional suppression that grows with $p_T$ \cite{Mehtar-Tani:2017web}.

\subsection{Color decoherence effects}

We are now equipped to write the medium-modified EEC evolution equation at LL accuracy, including the transition from color coherence to decoherence. The corresponding anomalous dimension is obtained from \eqn{eq:gamma-incoh} as follows:
\begin{align}\label{eq:gamma-coh}
\bgamma_\med(3)
\equiv
\begin{pmatrix}
\frac{25}{6}C_F \,S_{\rm dip}(\theta)  \quad -\frac{17}{15}n_f\\
 -\frac{7}{6}C_F \,S_{\rm dip}(\theta)\,  \quad \frac{14}{5}C_A S_{\rm dip}(\theta) +\frac{2}{3}n_f 
\end{pmatrix},
\end{align}
after substituting $S_g \to S_{\rm dip}(\theta)$, with the fully evolved subjet quenching factors replacing the bare ones in \eqn{eq:antenna-qw}, and using $S_q^2/S_g \sim 1$ in the large-$N_c$ limit.

Hence, 
\begin{align}\label{eq:EEC-sol-diff-med-2} 
   \frac{\del\bSigma(R/\chi) }{\del \ln \chi}  = \frac{\alpha_s( p_T\chi)}{2\pi} (1,1) \cdot \bgamma_{\rm med}(\chi) \cdot\, \exp\left[- \frac{\alpha_s}{4\pi} \int_{\chi}^R \frac{\rmd \theta^2}{\theta^2}\alpha_s( p_T\theta)\,\bgamma_{\rm med}(\theta) \right]\,.
\end{align}
Equations~\eqref{eq:EEC-sol-diff-med-2} and \eqref{eq:gamma-coh} are the main results of this work.

Turning to the full jet observable, the medium-modified jet EEC takes the form
\begin{align}
    \Sigma_{\rm jet}(R/\chi) =\frac{f_q \, \Qc_q(R) \, \Sigma^\med_q(R/\chi) +f_g  \,\Qc_g(R) \,\Sigma^\med_g(R/\chi)} {f_q  \,\Qc_q(R) +f_g  \,\Qc_g(R)}\,,
\end{align}
where the denominator is nothing but the jet nuclear modification factor 
\begin{align}
 R^{\rm jet}_{AA}(R)= f_q  \,\Qc_q(R) +f_g \, \Qc_g(R) \,.
\end{align}

We emphasize that these expressions do not account for soft medium-induced radiation or medium response, which are expected to become increasingly important at angles larger than $\theta_c$. Our approach can thus be viewed as a first QCD framework for the in-medium EEC that systematically resums color-decoherence effects in the hard-collinear sector, and may provide a basis for future extensions incorporating soft modes, for instance through matching onto an appropriate effective field theory.

To conclude this section, before turning to a phenomenological test of the framework, we can anticipate its qualitative implications. In the absence of coherence effects, the dipole quenching factor would remain close to its fully decoherent limit, $S_{\rm dip}\sim Q_q^2\ll1$, throughout the angular range inside the jet, leading to a strong modification of the EEC that is not supported by the data \cite{CMS:2025ydi}. Conversely, assuming fully coherent energy loss over the entire jet phase space is not necessarily justified either. A careful analysis of the data is therefore required to resolve the transition from coherent to decoherent energy loss, characterized by $S_{\rm dip}\to1$ for $\theta<\theta_c$ and a gradual departure from unity at larger angles.
\section{EEC phenomenology and medium resolution}\label{sec:eec_pheno}
In this section, we discuss the effects of the medium-modified anomalous dimension on the EEC in Eq.~\eqref{eq:EEC-sol-diff-med-2}. 
Before turning to heavy-ion collisions, however, we first construct the proton--proton ($pp$) reference using the solution in Eq.~\eqref{eq:EEC-sol-diff}, including running-coupling effects. To do so, we introduce an infrared regulator to phenomenologically describe the transition to the nonperturbative regime, which occurs when $p_T \chi\sim Q_{\rm N.P.}\sim 1~{\rm GeV}$.

\begin{figure}
\begin{center}
\includegraphics[width=0.6\textwidth]{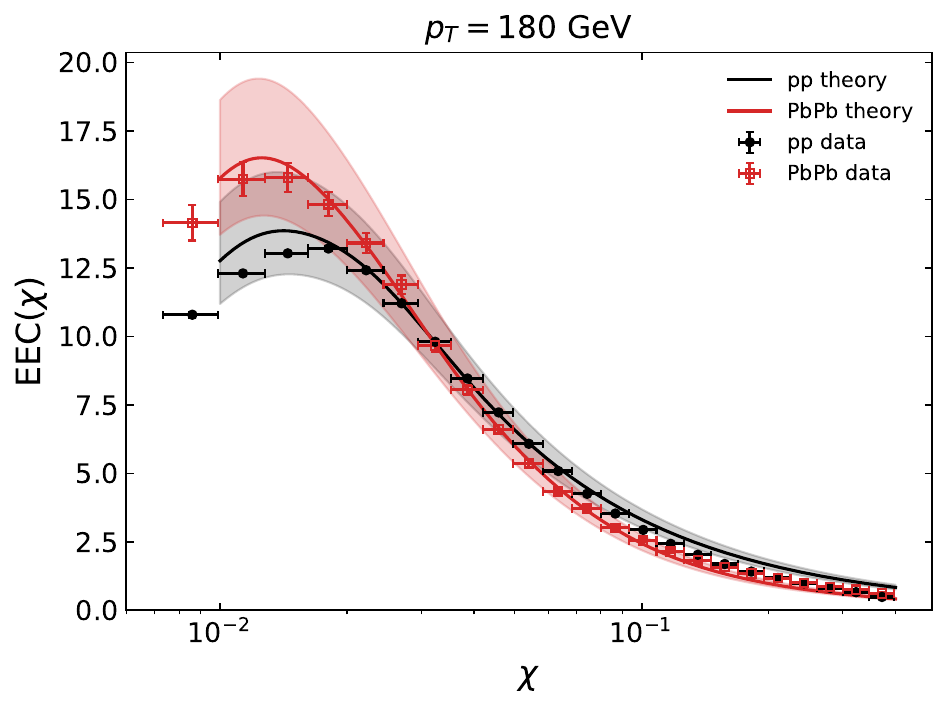} 
\end{center}
\caption{
EEC model calculation for $pp$ and PbPb collisions compared with CMS data at $p_T=180$--$200$ GeV and $R=0.4$. The parameters are obtained from a joint fit to the EEC and $R_{AA}$ data, yielding $\theta_c=0.06$, $S_q\simeq 0.65$, and $S_g\simeq 0.4$. The band represents the theoretical uncertainty estimated by varying the renormalization scale by a factor of two.  
 }
\label{fig:eec_pp_pbpb}
\end{figure}

Including the one-loop running of the coupling, the EEC takes the form
\begin{align}
\frac{\rmd {\bs \Sigma}}{\rmd \chi}
=
\frac{\alpha_s(p_T \chi)}{2\pi \chi}
\,(1,1)\cdot {\bs \gamma}(3)\cdot 
\left(
\frac{\alpha_s(p_T R)}
     {\alpha_s(p_T \chi)}
\right)^{\frac{{\bs \gamma}(3)}{\beta_0}}
\, ,
\end{align}
where $\beta_0=(11C_A-2n_f)/3$ is the one-loop coefficient of the QCD beta function.

We model the transition to the non-perturbative region by replacing the perturbative scale $\mu\equiv p_T \chi$ with a flavor-dependent infrared-regulated scale,
$\mu^2
\to 
\left(\mu^\beta+m_{q,g}^\beta\right)^{2/\beta}\,$. The infrared scales $m_q$ and $m_g$, together with the interpolation parameter $\beta$, are determined from the $pp$ data. In addition, we multiply the leading-logarithmic result by an overall $K$-factor which parametrizes corrections to the normalization not captured at LL accuracy. 

Taking the quark and gluon jet fractions from \textsc{Pythia} simulations (see Appendix~I) leaves four free parameters, $(\beta,K,m_q,m_g)$. We determine them through a simultaneous fit to the CMS EEC data in the four transverse-momentum intervals $100<p_T<120~{\rm GeV}$, $120<p_T<140~{\rm GeV}$, $140<p_T<160~{\rm GeV}$, and $160<p_T<180~{\rm GeV}$. The fit yields
$\beta=4.4\,$,
$K=2.6\,$,
$m_q=2.1~{\rm GeV}$\, and
$m_g=4.0~{\rm GeV}$\,. 

The explicit angular dependence of ${\bs \gamma}_{\rm med}(\theta)$ reflects the transition between coherent and decoherent energy loss.

The result of fitting \eqn{eq:EEC-sol-diff-med-2} to the vacuum and medium data is shown in figure~\ref{fig:eec_pp_pbpb}. The band accounts for a factor-of-two variation of the running-coupling scale and provides an estimate of higher-order corrections. The fit is performed jointly with the jet $R_{AA}$, allowing us to simultaneously extract the quark and gluon nuclear modification factors. We find values compatible with the expected Casimir scaling, $\Qc_g\sim (\Qc_q)^{C_A/C_F}$~\cite{Bossi:2026}, namely,
\beq
\Qc_q \sim 0.65, \quad \Qc_g \sim 0.4.
\eeq
By contrast, in the fully coherent limit, $\theta_c\to \infty$, the quality of the fit deteriorates and the extracted quenching factors are approximately $\Qc_q\sim 0.8$ and $\Qc_g\sim 0.2$, which are incompatible with Casimir scaling.

\begin{figure}
\begin{center}
\includegraphics[width=0.6\textwidth]{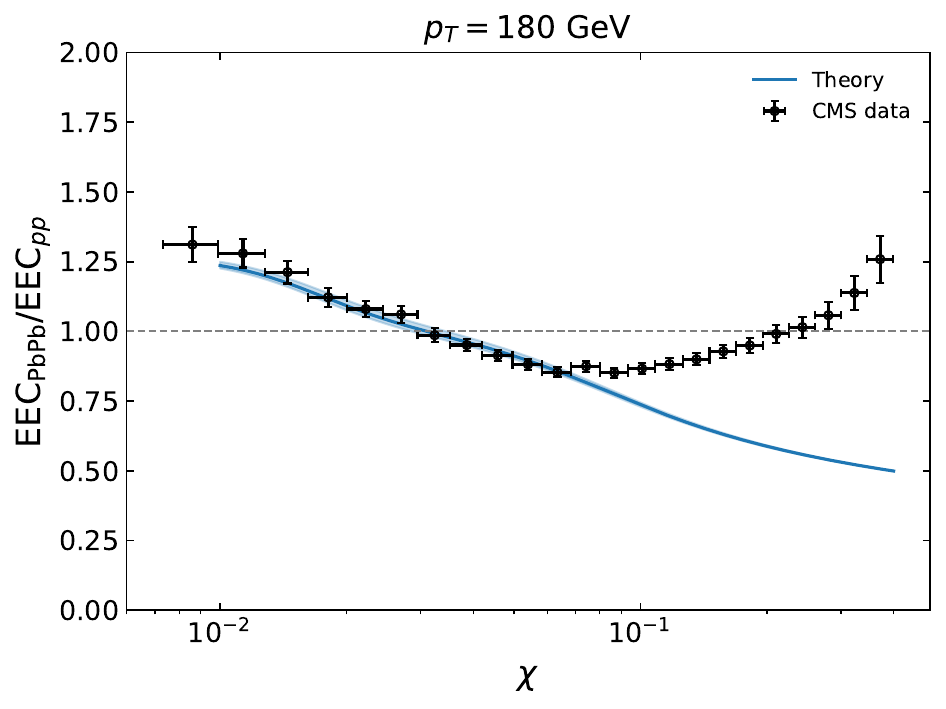} 
\end{center}
\caption{
The ratio of the EEC in PbPb to $pp$ collisions compared with CMS data at $p_T=180$--$200$ GeV and $R=0.4$. The band represents a factor-of-two variation of the renormalization scale, which largely cancels in the ratio.  
 }
\label{fig:eec_pp_pbpb_ratio}
\end{figure}

\begin{figure}
\begin{center}
\includegraphics[width=0.45\textwidth]{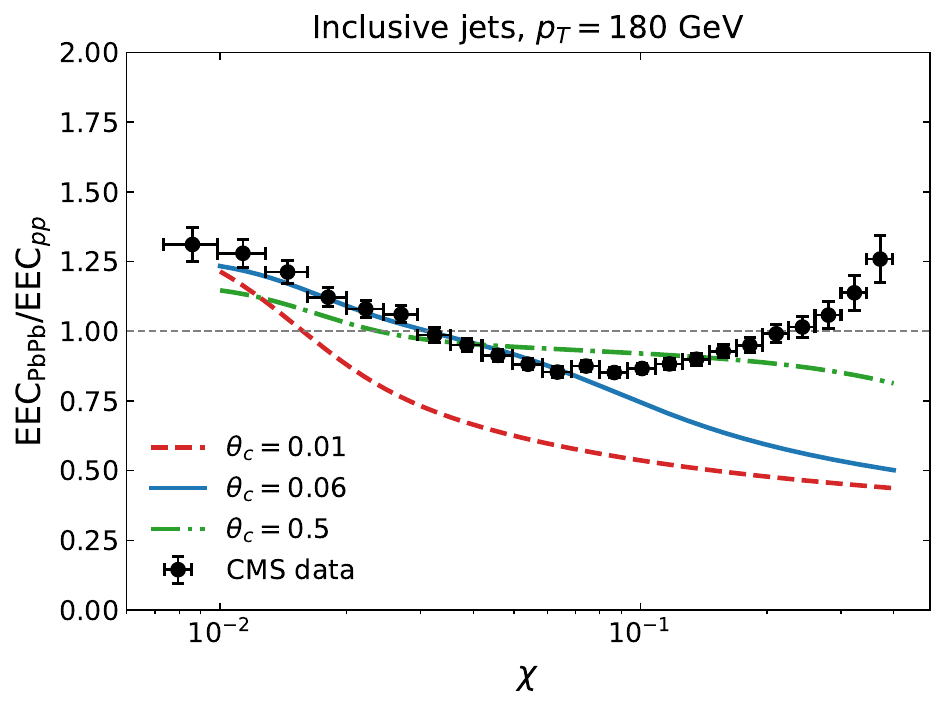} \includegraphics[width=0.45\textwidth]{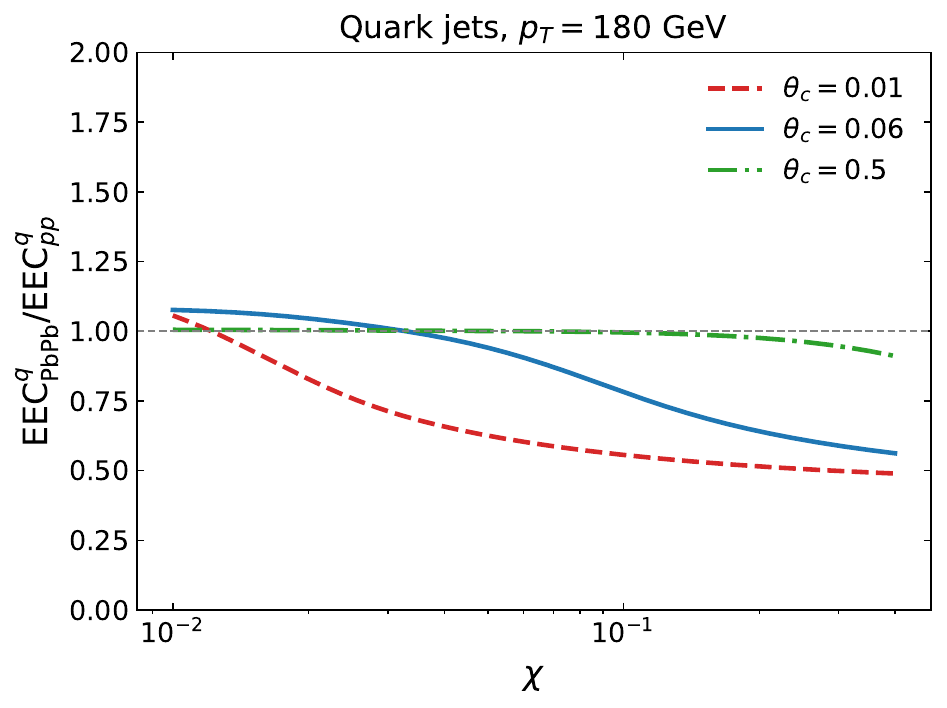} 
\end{center}
\caption{
Left: Ratio of the EEC in PbPb to $pp$ collisions compared with CMS data at $p_T=180$--$200$ GeV and $R=0.4$. Right: Corresponding ratio for quark-initiated jets, illustrating that in the coherent limit, $\theta_c \approx 0.5$, the quark EEC remains essentially unmodified, while the inclusive jet EEC (left) still exhibits a modification due to flavor-selection bias, arising from the stronger energy loss of gluon jets relative to quark jets. The variation with $\theta_c$ is shown for fixed quark and gluon quenching factors, extracted from a joint fit to the ATLAS $R_{AA}$ and CMS EEC data, which yields $\theta_c \simeq 0.06$. The data favor an intermediate scenario between the fully coherent and decoherent limits.
 }
\label{fig:eec_pp_pbpb_ratio_q}
\end{figure}

The ratio of the PbPb to pp EEC is shown in figure~\ref{fig:eec_pp_pbpb_ratio}. We see that our QCD evolution, which accounts only for hard-collinear energy loss, provides a good description of the small-angle region, $\chi<0.1$. The best fit is obtained for a resolution angle $\theta_c\sim 0.06$. This scale is well constrained by the data, since sizable variations of $\theta_c$ lead to substantial deviations, as illustrated in figure~\ref{fig:eec_pp_pbpb_ratio_q}. In this figure, we compare the best-fit value $\theta_c\sim 0.06$ with a more coherent scenario, $\theta_c\sim 0.5$, and a more decoherent scenario, $\theta_c\sim 0.01$, neither of which provides a satisfactory description of the data.

For reference, figure~\ref{fig:eec_pp_pbpb_ratio_q} also shows the corresponding ratio for the quark distribution alone. This removes the effect of the so-called selection bias, whereby the different energy loss of quark and gluon jets modifies the flavor composition of the selected jet sample and can therefore produce an apparent modification of the EEC without modifying the substructure of jets of a given flavor. The quark-only distribution makes the role of color decoherence particularly transparent. In the approximately coherent case, $\theta_c\sim 0.5$, the ratio remains close to unity, while for smaller $\theta_c$ the departure from unity directly reflects a genuine modification of the hard-collinear jet substructure induced by color decoherence.

At larger angles, additional physics becomes relevant and is commonly attributed to soft medium-induced radiation and/or medium response. Our analysis provides a natural starting point for investigating this soft component in future work, by coupling the evolution equations for the hard-collinear modes developed here to an effective description of the soft sector. Such a framework could be directly tested against the large-angle enhancement observed in the PbPb-to-pp ratio in figure~\ref{fig:eec_pp_pbpb_ratio}. This may provide the missing physics needed to understand the origin of this enhancement and, more generally, offer a systematic framework for addressing the questions surrounding its microscopic origin that have attracted considerable attention in recent years~\cite{Andres:2022ovj,Barata:2023bhh,Chen:2022muj}.

\section{Conclusions and outlook}
In this work, we have developed a leading-logarithmic (LL) framework for computing energy-energy correlators in vacuum and in the quark-gluon plasma using a generating-functional approach, focusing on hard-collinear modes at small angles. After recovering the known LL vacuum result, we derived evolution equations for the EEC in the medium. Combining these equations with the nonlinear DGLAP evolution governing the quenching factors in Laplace space reveals a simple structure for the medium modification: the vacuum anomalous dimension is dressed by the ratio of the quenching factors of the two daughter partons to that of their parent. This ratio encodes the relative suppression of resolved subjets in the parton cascade with respect to the total color charge of the jet. In this way, the evolution naturally separates the quenching of the total color charge from the additional suppression that arises as new color charges are resolved along the parton cascade.

We then showed how the transition from coherent to incoherent energy loss can be incorporated within this framework. At LL accuracy, strong angular ordering implies that at most one splitting at a time probes the transition region around the medium resolution angle $\theta_c$. In the large-$N_c$ limit, the resulting color structure further simplifies, and the medium modification of the anomalous dimension can be expressed entirely in terms of dipole quenching factors. The dipole factor therefore provides a direct link between the angular evolution of the EEC and the ability of the medium to resolve the internal color structure of the jet. This establishes the EEC as a particularly natural observable for resolving the transition between coherent energy loss at short angular scales and the independent quenching of resolved color charges at larger angles.

This formulation can be confronted with CMS EEC data through a joint analysis with the ATLAS jet $R_{AA}$ for $R=0.4$ jets. The two observables provide complementary information: while the inclusive jet suppression primarily constrains the overall magnitude and flavor dependence of the quenching, the angular dependence of the EEC is directly sensitive to the onset of color decoherence. We restrict the analysis to angles of order $0.1$ or smaller, where the hard-collinear dynamics considered here is expected to dominate, thereby avoiding the large-angle enhancement commonly attributed to soft physics such as medium-induced radiation and medium response. Within this region, the combined analysis allows us to extract a medium resolution angle of $\theta_c\approx 0.05$--$0.06$. Importantly, by employing the antenna energy-loss distribution derived in Ref.~\cite{Mehtar-Tani:2017ypq}, this extraction goes beyond the commonly used phenomenological modeling of the coherence-to-decoherence transition as a step function \cite{Caucal:2018ofz,Barata:2023bhh}. The continuous angular dependence of the antenna quenching factor retains sensitivity to the magnitude of the medium resolution scale, rather than merely distinguishing emissions above and below a prescribed angular boundary. A more detailed phenomenological analysis is presented in the companion Letter \cite{Bossi:2026}.

Interestingly, these results suggest that neither the fully coherent nor the fully incoherent limit provides an adequate description of the data, indicating genuine sensitivity to the transition between the two regimes. This opens the prospect of a quantitative extraction of the medium resolution scale and, ultimately, the jet-quenching parameter in a more comprehensive global analysis incorporating theoretical uncertainties and additional effects, such as the collision geometry.

More generally, this framework provides a basis for a systematic extraction of medium properties through combined analyses of observables with complementary sensitivities. An important future direction is to match the present hard-collinear description to an effective theory incorporating the in-cone medium-induced gluon cascade and, potentially, medium response \cite{Blaizot:2013vha,Mehtar-Tani:2024mvl}. Such an extension would make it possible to quantitatively address the large-angle enhancement of the EEC and exploit it to extract complementary information about medium-induced dynamics.

The present framework also provides a natural starting point for systematically incorporating higher-order corrections and extending the analysis to other jet-substructure observables. Particularly interesting in this respect are groomed observables, which suppress sensitivity to soft physics and may therefore provide cleaner access to the hard-collinear dynamics and color-coherence effects considered here. More broadly, extending this approach to a set of observables with different sensitivities to the flavor, angular structure, and soft components of jet energy loss could provide a path toward a multi-observable tomography of the quark-gluon plasma.

\vspace{0.5cm}
\begin{acknowledgments}
We thank Paul Caucal for insightful discussions on the theoretical aspects of this work, and Hannah Bossi and Maxence Larose for their collaboration on the phenomenological analysis presented in the companion Letter.  Y.~M.~T. was supported by the U.S. Department of Energy under Contract No. DE-SC0012704. 
\end{acknowledgments}

\bibliographystyle{utcaps}
\bibliography{main.bib}

\end{document}